\documentclass[prx,a4paper,aps,twocolumn,superscriptaddress,10pt,bibnotes]{revtex4-1}
\usepackage{graphicx,color,bbm}
\usepackage{dcolumn}
\usepackage{bm}
\usepackage{amsmath,amssymb,mathrsfs}
\usepackage{url}
\usepackage{hyperref}

\setcitestyle{sort&compress,numbers}

\def\>{\rangle}
\def\<{\langle}

\newcommand{\map}[1]{\mathcal{#1}}

\begin{document}

\title{Demonstration of superior communication through thermodynamically free channels in an optical quantum switch}

\author{Hao Tang}
\thanks{These two authors contributed equally.}
\affiliation{CAS Key Laboratory of Quantum Information, University of Science and Technology of China, Hefei, 230026, China}
\affiliation{CAS Center For Excellence in Quantum Information and Quantum Physics, University of Science and Technology of China, Hefei, 230026, China}

\author{Yu Guo}
    \thanks{These two authors contributed equally.}
    \affiliation{CAS Key Laboratory of Quantum Information, University of Science and Technology of China, Hefei, 230026, China}
    \affiliation{CAS Center For Excellence in Quantum Information and Quantum Physics, University of Science and Technology of China, Hefei, 230026, China}
    
\author{Xiao-Min Hu}
\affiliation{CAS Key Laboratory of Quantum Information, University of Science and Technology of China, Hefei, 230026, China}
\affiliation{CAS Center For Excellence in Quantum Information and Quantum Physics, University of Science and Technology of China, Hefei, 230026, China}
\affiliation{Hefei National Laboratory, University of Science and Technology of China, Hefei, 230088, China}

\author{Yun-Feng Huang}
\affiliation{CAS Key Laboratory of Quantum Information, University of Science and Technology of China, Hefei, 230026, China}
\affiliation{CAS Center For Excellence in Quantum Information and Quantum Physics, University of Science and Technology of China, Hefei, 230026, China}
\affiliation{Hefei National Laboratory, University of Science and Technology of China, Hefei, 230088, China}

\author{Bi-Heng Liu}
\email{bhliu@ustc.edu.cn}
\affiliation{CAS Key Laboratory of Quantum Information, University of Science and Technology of China, Hefei, 230026, China}
\affiliation{CAS Center For Excellence in Quantum Information and Quantum Physics, University of Science and Technology of China, Hefei, 230026, China}
\affiliation{Hefei National Laboratory, University of Science and Technology of China, Hefei, 230088, China}

\author{Chuan-Feng Li}
\affiliation{CAS Key Laboratory of Quantum Information, University of Science and Technology of China, Hefei, 230026, China}
\affiliation{CAS Center For Excellence in Quantum Information and Quantum Physics, University of Science and Technology of China, Hefei, 230026, China}
\affiliation{Hefei National Laboratory, University of Science and Technology of China, Hefei, 230088, China}

\author{Guang-Can Guo}
\affiliation{CAS Key Laboratory of Quantum Information, University of Science and Technology of China, Hefei, 230026, China}
\affiliation{CAS Center For Excellence in Quantum Information and Quantum Physics, University of Science and Technology of China, Hefei, 230026, China}
\affiliation{Hefei National Laboratory, University of Science and Technology of China, Hefei, 230088, China}

\begin{abstract}
The release of causal structure of physical events from a well-defined order to an indefinite one stimulates remarkable enhancements in various quantum information tasks. Some of these advantages, however, are questioned for the ambiguous role of the control system in the quantum switch that is an experimentally realized process with indefinite causal structure. In communications, for example, not only the superposition of alternative causal orders, but also the superposition of alternative trajectories can accelerate information transmissions. Here, we follow the proposal of Liu et al. [Phys. Rev. Lett. \textbf{129}, 230604 (2022)], and examine the information enhancement effect of indefinite causal orders with the toolkit of thermodynamics in a photonic platform. Specifically, we simulate the thermal interaction between a system qubit and two heat baths embedded in a quantum switch by implementing the corresponding switched thermal channels. Although its action on the system qubit only is thermally free, our results suggest that the quantum switch should be seen as a resource when the control qubit is also considered. Moreover, we characterize the non-Markovian property in this scenario by measuring the information backflows from the heat baths to the system qubit.

\end{abstract}

\maketitle

\textit{Introduction.---} 
Over the last decades, quantum information processing has been extended to scenarios where basic quantum operations are executed in a superposition of causal orders\cite{chiribella2012perfect,chiribella2013quantumcomputations,oreshkov2012quantum}. Such exotic processes with indefinite causal orders (ICOs) were originally proposed by Hardy in the context of reconciling quantum mechanics and general relativity \cite{hardy2007towards,hardy2009quantum} and have attracted great interests in the community. It has been recognized that ICO offers remarkable advantages in fields such as channel discrimination \cite{Procopio2015Experimental}, communication complexity \cite{Guerin2016communicationComplexity,wei2019experimentalCommunication}, communication \cite{Ebler2018EnhancedCommunication,Salek2018QuantumCommunication,Chiribella2021perfectQuantumCommunication,procopio2020sending,procopio2019communication,Sazim2021ClassicalCommunication,guo2020experimental,goswami2020IncreasingCommunication,Rubino2021Communication}, computation \cite{Araujo2014ComputationalAdvantage,renner2022advantage,taddei2021computational,liu2023experimentally}, thermodynamics \cite{felce2020QuantumRefrigeration,Guha2020Thermodynamic,Simonov2022WorkExtraction,chen2021indefinite,nei2022NMRswitch,cao2022quantumSimulation,zhu2023prl}, metrology \cite{zhao2019QuantumMetrology,chapeau2021noisy,kurdzialek2023using,liu2023optimal,yin2023experimental}, and others\cite{quintino2019reversing,bavaresco2021strict,trillo2205universal,Schiansky22TimeReversal,koudia2023deterministic,zuo2023coherent,gao2023measuring}; see Ref. \cite{rozema2024experimental} for a review.

In communications, the advantages of ICO have been reported in a series of theoretical \cite{Ebler2018EnhancedCommunication,Salek2018QuantumCommunication,Chiribella2021perfectQuantumCommunication} and experimental \cite{guo2020experimental,goswami2020IncreasingCommunication} works. Specifically, if two noisy channels are used in a fixed causal structure, one obtains a noisier channel. However, if they are embedded in a quantum switch\cite{chiribella2012perfect,chiribella2013quantumcomputations,oreshkov2012quantum}, the overall noise can actually decrease. This can benefit the transmission of both classical \cite{Ebler2018EnhancedCommunication,goswami2020IncreasingCommunication} and quantum \cite{Salek2018QuantumCommunication,Chiribella2021perfectQuantumCommunication,guo2020experimental} information.
However, there are debates over to what extent the advantages mentioned above are specific to ICO. Similar noise reduction effects have been reported in proposals that only involve the superposition of channels \cite{Abbott2020coherentControl,Rubino2021Communication,pang2023experimental,Guerin2019quantumControlledNoise}. 
Although there are still discussions regarding this effect, it is believed the existence of a noiseless control system in the quantum switch plays an ambiguous role, since it is involved in the decoding operation and partial information might bypass the noise acting on the target space. Analogous effects have also been reported in ICO-assisted thermodynamics \cite{Capela2023ReassessingThermodynamic} and metrology \cite{mothe2023reassessing}.

Recently, Liu et al. \cite{PhysRevLett.129.230604} investigated ICO-enhanced communication from the quantum thermodynamics point of view. Basically, there is a strong connection between information theory and thermodynamics via the central role of entropy in both fields. The direction of dynamical evolution in nature is governed by the laws of thermodynamics. In the quantum realm, the accuracy of any information task associated to particular quantum evolution can be bounded by the thermal resource that is consumed in the task \cite{chiribella2022nonequilibrium}. In Ref. \cite{PhysRevLett.129.230604}, an upper bound for the mutual information was provided for the case when information is transmitted through two thermal channels embedded in a quantum switch. In this scenario, the thermal channels obey the first and second laws of thermodynamics and thus are thermally free operations. The switched channel, however, is free only when the target system is considered and becomes a kind of thermal resource when both the target and control systems are involved. The increases in information capacity consume the free energy of coherence, a kind of thermal resource, of the switch's control qubit.

Here we report the experimental demonstration of superior communication through two thermodynamically free channels embedded in ICO\cite{PhysRevLett.129.230604}. To achieve this, we construct an optical quantum switch and the thermal channels to simulate the scenario in which the thermocontacts between a single qubit and the heat baths occur in a superposition of causal orders. Notably, the thermal channels we used obey both the first and second laws of thermodynamics and thus are a kind of free operation in the quantum resource theory (QRT) framework \cite{chitambar2019quantum}. Our results show an increase in communication rate and a decrease in free energy of coherence of the control qubit occur at the same time, suggesting the switch is a resource in this scenario. In addition, we observe a phenomenon of information backflow, a typical feature of non-Markovianinty\cite{breuer2002theory,liu2011experimental,OQDrmp}, from the heat bath to the system qubit during their thermal interaction, which only appears in the existence of ICO in this protocol.

\begin{figure}[tb]      
    \begin{center}
    \includegraphics [width=0.7\columnwidth]{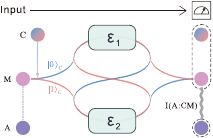}
    \end{center}
    \caption{\emph{Process schematic.} Qubits $M$ and $A$ are prepared in a separable but correlated state. $M$ undergoes a switched thermal channel to simulate the scenario where it interacts with two heat baths that are in a superposition of causal orders. The quantum mutual information $I(A:CM)$ between $A$ and $CM$ is analyzed, where $C$ is the control qubit of the quantum switch.
    }
    \label{fig:concept}
\end{figure}

\textit{Theoretical protocol.---} We start with a brief review of communication through two thermodynamically free channels, $\map E$ and $\map F$, that act in a superposition of causal orders. In the standard setting, the causal relations of the two channels are well-defined, and so that the information carrier encounters $\map E$ before $\map F$, or vice versa. The overall channel in either configurations is $\map E \circ \map F$ or $\map F \circ \map E$. In the quantum realm, the two channels can be arranged in a quantum superposition of these two configurations by associating them to two orthogonal state $|0\>$ and $|1\>$ of a control qubit. In this case, the information carrier undergoes $\map E \circ \map F$ and $\map F \circ \map E$ simultaneously and the causal relation of them can be described by the quantum switch transformation \cite{chiribella2012perfect,chiribella2013quantumcomputations,oreshkov2012quantum}, $\map S_\omega  :  (\map E, \map F) \mapsto \map S_\omega (\map E, \map F)$, defined as
\begin{align}\label{switchdef1}
	\map S_{\omega} (\map E, \map F)\left(\rho\right):= \sum_{i,j} W_{ij} \left(\rho \otimes \omega \right) W_{ij}^\dagger,  \end{align}
with $\rho$ and $\omega$ being the density matrices of the information carrier and the control qubit and
\begin{align}\label{switchdef2}
	W_{ij}:=E_i F_j \otimes |0\>\<0| + F_j  E_i \otimes   |1\>\<1|  \,,
\end{align}
where $\{E_i\}$ and $\{F_j\}$ are the Kraus operators of $\map E$ and $\map F$. 

It is crucial for the protocol to ensure that the channels used in the quantum switch are thermodynamically free, that is, they are free operations in QRT for quantum thermodynamics\cite{sparaciari2020first,brandao2015second,masanes2017general}. In the QRT of \emph{athermality} \cite{yunger2016beyond}, a well-studied QRT for quantum thermodynamics, the free operation is identified as those that obey the first law of total energy conservation during the thermal interaction between a system $S$ and a heat bath $B$. Here, the heat bath with temperature $T$ is assumed to be large enough and is in free state, namely in the Gibbs thermal state $\tau_T=exp(-H_{B}/kT)/Z$ with $H_B$ being its Hamiltonian, $k$ Boltzmann's constant, and $Z=Tr[exp(-H_{B}/kT)]$ a normalizing partition function. The interaction is required to be some unitary $U$ that commutes with the total Hamiltonian, i.e. $[U, H_S\otimes \mathbbm{I}+\mathbbm{I}\otimes H_B]=0$, where $H_S$ is the Hamiltonian of the system and $\mathbbm{I}$ is the identity in corresponding Hilbert space. And the free operation, the so-called \emph{thermal operation}, can then be constructed by tracing over an arbitrary subsystem of the system and heat bath after the interaction, yielding a completely positive and trace preserving (CPTP) channel $\varepsilon$ from the original system to the remaining subsystem after the discarding. In the case when the discarded subsystem is exactly the heat bath, it can be proven that such a thermal operation acts invariantly on the Gibbs thermal state, i.e. $\varepsilon(\tau_T)=\tau_T$ \cite{PhysRevLett.129.230604}. Thus, the channel acting on the original system, referred to as \emph{thermal channel} in the following, is Gibbs-preserving with respect to the Hamiltonian $H_S$, obeying the second law of free energy non-increase.

In the simplest nontrivial case, the protocol in Ref. \cite{PhysRevLett.129.230604} involves a pair of qubits, $M$ and $A$, and two identical heat baths in a quantum switch (see Fig.~\ref{fig:concept} where $C$ is the control qubit of the switch). The Hamiltonian for qubit $M$ ($A$) is chosen to be $H_{M(A)}=\Omega(|e\>\<e|^{M(A)}-|g\>\<g|^{M(A)})$ with $\{|e\>^{M(A)}, |g\>^{M(A)}\}$ being the energy basis and $\Omega$ the energy gap between them (we set $\Omega=1$ for simplicity in the following). For the interaction between a system qubit and a heat bath, the exact form of the thermal channel $\varepsilon$ that satisfies energy-preserving and Gibbs-preserving conditions can be specified according to the collision model \cite{PhysRevLett.88.097905}. In this model, the heat bath is assumed to consist of infinite qubits (the Hamiltonian is  $H=\sum_{i=1}^\infty H_{i}=\sum_{i=1}^\infty(|e\>\<e|^i-|g\>\<g|^i)$) and the system qubit collides only once with a single qubit from the heat bath through a unitary operation. Then the channel $\varepsilon$ can be determined to be a partial swap $U=\sqrt{1-s^2}\mathbbm{I}+isSWAP$ between the system qubit and one of the heat bath qubit, where $s\in [0, 1]$ is the thermalization strength (see the Supplemental Materials \cite{SM} for more details). 

\begin{figure*}[!ht]      
    \begin{center}
    \includegraphics [width=1.4\columnwidth]{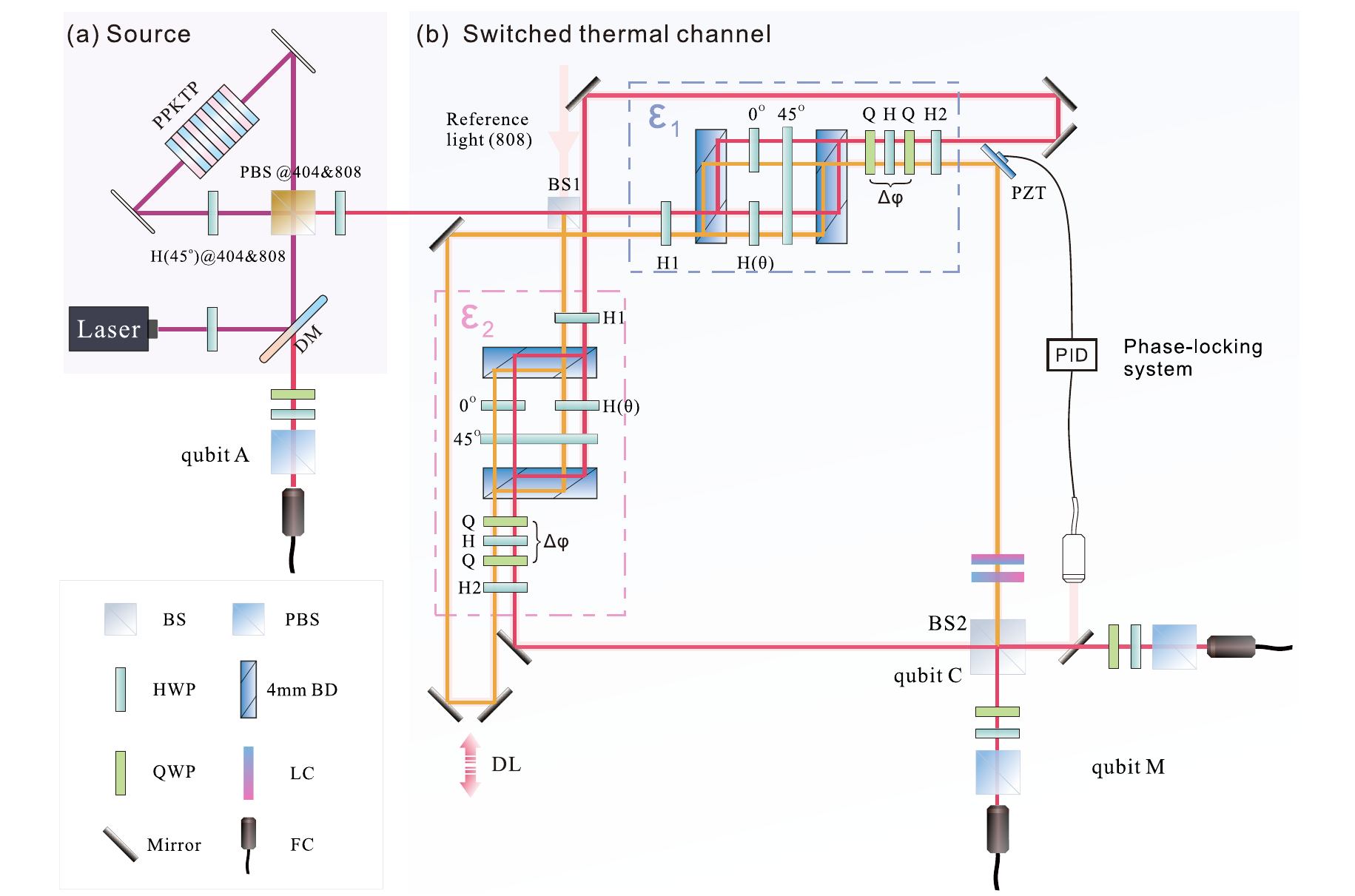}
    \end{center}
    \caption{\emph{The experimental setup.}
    (a) Initial input states preparation. A 404 nm continuous wave laser pumps a type-II cut ppKTP crystal, generating photon pairs at 808~nm. The photons are used to encode qubits $A$ and $M$ in polarization. (b) Switched thermal channel construction. A beam splitter (BS1) introduces two path modes to encode the control qubit $C$, and a Mach-Zehnder interferometer constitutes the optical quantum switch. The two thermodynamically free channels are shown in the blue and red dashed boxes, respectively.
    Measurement of qubit $C$ is accomplished by BS2 and a LC, while measurement of qubits $A$ and $M$ is performed with a QWP, a HWP, and a PBS.
    The phase-locking system consists of a reference laser at 808~nm, a photon detector, a piezo-transducer (PZT), and a PID regulator.
    BS: beam splitter; PBS: polarizing beam splitter; HWP: half-wave plate; BD: beam displacers; QWP: quarter-wave plate; LC: liquid crystal phase plate; FC: fiber coupler; DL: trombone-arm delay line.
    }
    \label{fig:Setup}
\end{figure*}

The protocol can then be sketched in three steps. Firstly, $M$ carries information about $A$ by being prepared in the state,
\begin{align}\label{inputstate}
	\rho_{AM}^{in}=p|g\rangle \langle g|_{A}\otimes|g\rangle \langle g|_{M}+(1-p)|e\rangle \langle e|_{A}\otimes|e\rangle \langle e|_{M},
\end{align}
which is uncorrelated with the heat baths and $p\in [0,1]$. Then, $M$ undergoes thermocontact with the switched heat baths, which act as a noisy channel $\map S_{\omega} (\varepsilon_{1}, \varepsilon_{2})$ impressed on $M$. Finally, the quantum mutual information $I(A:CM)$ between the output states of $A$ and $CM$ is analyzed to characterize the information that can be transmitted. Here, $I(X:Y)$ is defined as $I(X:Y)=S(\rho_X)+S(\rho_Y)-S(\rho_{XY})$, where $\rho_x$ is the process matrix of a system $x$ and $S(\rho)=-Tr(\rho \log_2{\rho})$ is the von Neumann entropy of $\rho$. Note that the maximum quantum mutual information over all input separable states is equivalent to the Holevo capacity of a quantum channel\cite{wilde2013quantum}.



\textit{Experimental setup.---} As illustrated in Fig.~\ref{fig:Setup}, we experimentally demonstrate communication through the switched thermal channel $\map S_{\omega} (\varepsilon_{1}, \varepsilon_{2})$ on a photonic system. A photon pair at 808~nm, generated through the spontaneous parametric down-conversion (SPDC) process in a type-II cut ppKTP crystal, serves as qubits $A$ and $M$ respectively. We utilize photon polarization to encode the basis of $A$ and $M$ where horizontal and vertical polarizations represent states $|g\>$ and $|e\>$ respectively. The mixed state described by Eq.~(\ref{inputstate}) is achieved by mixing its pure components with the probabilities $p$ and $1-p$.

The channel $\map S_{\omega} (\varepsilon_{1}, \varepsilon_{2})$ is composed by a control qubit $C$ and two  thermodynamically free channels $\varepsilon_{1,2}$. Qubit $C$ is encoded with the two spatial modes introduced by a beam splitter (BS1 in Fig.~\ref{fig:Setup}) and is initialized in the superposition state $|+\rangle_{C}= (|0\rangle_{C} + |1\rangle_{C})/\sqrt{2}$, representing that the switch is turned on. Depending on the state of qubit $C$ ($|0\rangle_{C}$ or $|1\rangle_{C}$), qubit $M$ undergoes the causal order  $\varepsilon_{2}\circ \varepsilon_{1}$ or vice versa. As we already know, the action of a quantum channel on a given quantum state of the system can be expressed in the Kraus representation. In our experiment, the Kraus operators of the thermal channels $\varepsilon_{1,2}$ are implemented with an assembly that is made up of two beam displacers (BDs), several half wave plates (HWPs), and two quarter wave plates (QWPs). Here, the two BDs are used to separate and recombine polarization modes of $M$, and the HWPs between the BDs control the amplitude of $M$. Additionally, two QWPs set at $45^\circ$, with a rotatable HWP sandwiched between them, can introduce any phase $\Delta\phi$ between horizontal and vertical polarizations. Details on the exact form of the Kraus operators of the thermal channels and their implementation are given in the Supplemental Materials ~\cite{SM}. 

\begin{figure*}[ht]      
    \begin{center}
    \includegraphics [width=1.4\columnwidth]{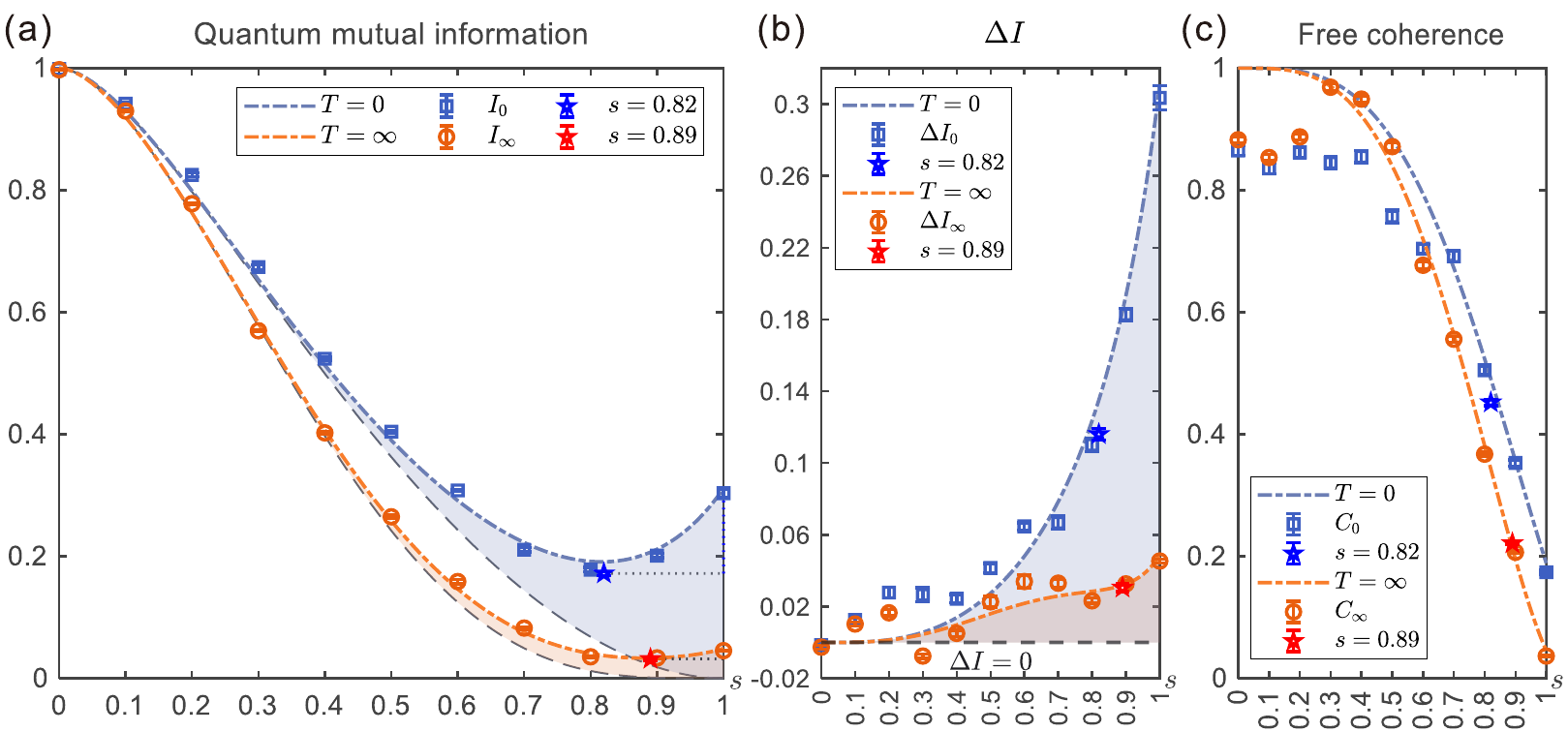}
    \end{center}
    \caption{\emph{The experimental results on (a) the quantum mutual information $I(A:CM)$, (b) the differences of quantum mutual information $\Delta I$, and (c) the free coherence $\mathcal{A}_C$ of $C$.} The blue squares and orange dots represent results for heat baths at $T=0$ and $T=\infty$, respectively. The dash-dot lines are theoretical curves when the switch is on, while the two gray dashed lines are the ones when the switch is off. The stars represent the results for the turning points. The increases in $\Delta I$ (b) and the decreases in the free coherence $\mathcal{A}_C$ (c) occur simultaneously. Error bars in this paper are estimated by Monte Carlo simulations.
    }
    \label{fig:Mutualinfo}
\end{figure*}

To determine the mutual information between $A$ and $CM$, we need to reconstruct the joint state $\rho_{ACM}$ through quantum state tomography \cite{PhysRevA.64.052312}, which requires a set of information-complete measurements on every qubit. In our experiment, the measurement on $C$ is accomplished using a beam splitter (BS2 in Fig.~\ref{fig:Setup}) and two orthogonally placed liquid crystal phase plates (LCs), while $M$ and $A$ are measured using a combination of a QWP, a HWP, and a polarization beam splitter (PBS). To ensure the phase stability of the Mach-Zehnder interferometer in the quantum switch (between BS1 and BS2), we adopt a phase-locking system that consists of a reference laser, a piezo-transducer, and a PID regulator. With the help of this system, we observe an overall visibility of 0.996 during 2 hours. See the Supplemental Materials ~\cite{SM} for details.

\textit{Experimental results.---} In our experiment, we test two switched thermal channels corresponding to the cases of two identical heat baths with zero ($T=0$) or infinite ($T=\infty$) temperature in the quantum switch. As mentioned above, the action of the switched heat baths on qubit $M$ is a Gibbs-preserving CPTP map, that is $\map S_{\omega} (\varepsilon_{1}, \varepsilon_{2})(\tau_{M})=\tau_{M}$ for arbitrary thermalization strength $s$. The Gibbs thermal states of $M$ with respect to $T=0$ and $T=\infty$ are $\tau_0=|g\rangle \langle g|$ and $\tau_\infty=\frac{1}{2}|g\rangle\langle g|+\frac{1}{2}|e\rangle \langle e|$ respectively. To check the Gibbs-preserving property, we prepare these thermal states, reconstruct their density matrices after the quantum switch, and calculate the fidelity of the reconstructed states with respect to the corresponding thermal states. The worst case fidelity exceeds $0.997$, thereby showing the high quality of our setup (see Ref.~\cite{SM} for details). 

Next, we measure the quantum mutual information of the switched channels. The initial state is chosen to be the one in Eq.~(\ref{inputstate}) with $p=0.5$. When $T=\infty$, this is the optimal input state to obtain the maximum $I(A:CM)$ for arbitrary thermal strength $s$. Thus, the quantum mutual information obtained in our experiment is exactly the Holevo capacity when $T=\infty$, while it serves as the permitted information transmission rate with respect to the particular input state when $T=0$. Our results of $I(A:CM)$ for $T=0$ and $T=\infty$ are illustrated in Fig.~\ref{fig:Mutualinfo}(a) as a function of $s$. Our experimental data are marked with blue squares ($T=0$) and orange dots ($T=\infty$), matching well with their theoretical predictions that are given as the dash-dotted lines. As a contrast, we also plot $I(A:CM)$ for the case when the switch is turned off in gray dashed lines by preparing the control qubit $C$ in $|0\rangle\langle 0|$. As shown, there are significant gaps in $I(A:CM)$ between the cases when the switch is on and off for both temperatures. These results verified the existence of the advantage of ICO in communication through thermodynamically three channels. We further show this advantage by plotting the increase in the quantum mutual information, labeled as $\Delta I$, in Fig.~\ref{fig:Mutualinfo}(b). It shows that $\Delta I$ is positively related to the thermal strength $s$. 

Although not being used to encode information, the control qubit $C$ is used to assist the decoding, e.g., during the calculation of $I(A:CM)$ in our case. In fact, it is pointed out that the action of the switched thermal channel on $C$ and $M$ as a whole needs to be treated as a thermal resource for the Gibbs-preserving property does not apply any more in this case \cite{PhysRevLett.129.230604}. To investigate the role of $C$ in this process, we calculate the free coherence \cite{lostaglio2015description} of $C$, given by $\mathcal{A}_C(\rho)=Tr[\rho(\log_2 \rho -\log_2 \mathcal{D}_H(\rho))]$ after the quantum switch. Here $\mathcal{D}_H$ is the operation that removes all coherence between energy eigenspaces. The free coherence can be seen as the normalized free energy of coherence $F_{coh}(\rho)=kT\mathcal{A}(\rho)$ \cite{PhysRevLett.129.230604}, which is closely related to the Gibbs free energy by $F_{coh}(\rho)=F(\rho)-F(\mathcal{D}_H(\rho))$ with $F(\rho)=Tr(\rho H)-kTS(\rho)$being the Gibbs free energy and $F(\mathcal{D}_H(\rho))$ being the classical free energy. The results, illustrated in Fig.~\ref{fig:Mutualinfo}(c), show a remarkable decrease in the free coherence of $C$, indicating the communication advantage of ICO in this protocol comes at a cost of the consumption of the thermal resource of quantum switch.
 
In addition, we observe a phenomenon of information backflow, a typical non-Markovian behaviour in open quantum dynamics\cite{breuer2002theory,liu2011experimental,OQDrmp}, during the thermocontact between qubit $B$ and the switched heat baths. The heat baths can be modeled as the environment that is difficult to control. When the switch is turned off, the information of system, i.e., $I(A:CM)$, dissipates until disappear into the heat bath as the thermalization proceeds. However, in the existence of ICO, there is a turning point (TP) of $s$, after which $I(A:CM)$ flows back from the switched heat baths into the system. The TPs are calculated to be $s=0.82$ and $s=0.89$ for $T=0$ and $T=\infty$ respectively. To characterize the non-Markovianinty $\mathcal{N}$, we adopt the measure in Ref.~\cite{luo2012quantifying}, that is,
\begin{align}\label{non-M}
    \mathcal{N}:=\int_{s_{TP}}^{s=1} \dfrac{d}{ds} I ds.
\end{align}
We examine $I(A:CM)$ at the TPs (stars in Fig.~\ref{fig:Mutualinfo}) and the differences of $I(A:CM)$ between $s=1$ and the TPs provide lower bounds for $\mathcal{N}$. In our experiment, the values of $\mathcal{N}$ are $0.132\pm 0.008$ and $0.014\pm 0.002$ for the heat baths at $T=0$ and $T=\infty$.

\textit{Conclusion.---} We have experimentally demonstrated enhanced communication through thermodynamically free channels in an optical quantum switch. The channel we constructed can be used to simulate the thermocontect between a qubit and heat baths that are embedded in a superposition of causal orders. Our experiment results suggest that the quantum switch should be seen as a resource rather than a free operation when the control qubit of the switch is taken into consideration. In this scenario, we also observed a quantum memory effect in the existence of ICO. Our work, together with Refs.~\cite{PhysRevLett.129.230604,xi2023experimental}, will stimulate similar analyses of the advantages of ICO on other tasks~\cite{Procopio2015Experimental,Guerin2016communicationComplexity,wei2019experimentalCommunication,Ebler2018EnhancedCommunication,Salek2018QuantumCommunication,Chiribella2021perfectQuantumCommunication,procopio2020sending,procopio2019communication,Sazim2021ClassicalCommunication,guo2020experimental,goswami2020IncreasingCommunication,Rubino2021Communication,Araujo2014ComputationalAdvantage,renner2022advantage,taddei2021computational,liu2023experimentally,felce2020QuantumRefrigeration,Guha2020Thermodynamic,Simonov2022WorkExtraction,chen2021indefinite,nei2022NMRswitch,cao2022quantumSimulation,zhu2023prl,zhao2019QuantumMetrology,chapeau2021noisy,kurdzialek2023using,liu2023optimal,yin2023experimental,quintino2019reversing,bavaresco2021strict,trillo2205universal,Schiansky22TimeReversal,koudia2023deterministic,zuo2023coherent,gao2023measuring}, which might also be done within QRT for thermodynamics.

\begin{acknowledgements}
We thank Ning-Ning Wang for valuable discussions. This work was supported by NSFC (No.~12374338, No.~11904357, No.~12174367,  No.~12204458, and No. 17326616), the Innovation Program for Quantum Science and Technology (No. 2021ZD0301200), the Fundamental Research Funds for the Central Universities, China Postdoctoral Science Foundation (2021M700138), China Postdoctoral for Innovative Talents (BX2021289). This work was partially carried out at the USTC Center for Micro and Nanoscale Research and Fabrication.

\textit{Note added.--} During the completion of our manuscript, we became aware of a work by Xi et al.\cite{xi2023experimental}, which independently demonstrated the protocol on a NMR platform and was concurrently posted to arXiv with ours.

\end{acknowledgements}

\bibliography{references}

\end{document}